\documentclass[twocolumn,aps, prb]{revtex4}

\usepackage{graphics}
\usepackage[pdftex]{graphicx}
\usepackage{dcolumn}
\usepackage{bm}

\newcommand{\de}{\partial}
\newcommand{\sech}[1]{\textrm{sech}\left(  #1\right)}
\newcommand{\eq}[2]{\begin{equation} \label{#1} #2 \end{equation}}
\newcommand{\eps}{\epsilon}

\begin{document}


\title{Multiple hydrodynamical shocks induced by Raman effect in photonic crystal fibres}
\author{C. Conti$^1$, S. Stark$^2$, P. St.J. Russell$^2$, F. Biancalana$^2$}

\affiliation{$^1$CNR-ISC, Department of Physics, University La Sapienza, P.le A.Moro 2, 00185 Rome, Italy \\
$^2$ Max Planck Institute for the Science of Light, 91058 Erlangen, Germany}
\date{\today}

\begin{abstract}
We theoretically predict the occurrence of multiple hydrodynamical-like shock phenomena in 
the propagation of ultrashort intense pulses in a suitably engineered photonic crystal fiber. 
The shocks are due to the Raman effect, which acts as a nonlocal term favoring their generation
in the focusing regime. It is shown that the problem is mapped to shock formation in the presence of a slope and a gravity-like
potential. The signature of multiple shocks in XFROG signals is unveiled.
\end{abstract}


\maketitle
\section{Introduction}
The existence of regimes where nonlinear optical propagation and Bose-Einstein condensation (BEC) 
obey hydrodynamical-like models is well known  \cite{Gurevich87, Anderson92, stringari96, Kodama99,Forest99}.
Among the resulting hydrodynamic-like phenomena for intense light pulses or spatial beams,
the most fascinating is certainly the formation of shocks \cite{LandauBookFluid}, which can also be considered in Bose-Einstein condensates (BEC) \cite{Rothenberg89, Cai97, ablowitz09, perezgarcia04, Kamchatnov04, Wan07, hoefer06,hoefer08a}.

Theoretical and experimental work on optical shock formation has been recently reported for nonlocal media,
like thermal liquids \cite{ghofraniha07,fratalocchi08,conti09}, liquid crystals \cite{Assanto08shock}
and photorefractive materials \cite{el07,fleisher10}.
In the time domain, shocks have been considered in connection with the carrier wave \cite{Flesch96} and quadratic media \cite{baronio06}. Spatiotemporal hyperbolic shocks have also been studied  \cite{berge02}, together with
related nonlinear X-wave generation \cite{conti03,bragheri07} and multidimensional effects \cite{armaroli09,hoefer09}.

When considering temporal shocks, if it is true that the development of extreme nonlinear optics \cite{Brabec00} unavoidably requires the consideration of 
ultra-wide spectral-band and intense excitations (common to all shock phenomena),
the presence of higher order effects radically 
affects shock formation and related regularization processes, such as undular bores \cite{WhithamBook, Gurevich73}.
In this respect, microstructured photonic-crystal fibers (PCFs) \cite{russell} offer an unprecedented framework for exploring the hydrodynamical properties of light. Indeed, not only do solid-core PCFs exhibit very pronounced nonlinear effects (mainly thanks to their small effective modal areas), but their dispersion profile can be engineered to a large degree. 
The latter circumstance is particularly
appealing for the physics of shock wave generation, because most of the dynamics described by nonlinear Schr\"odinger (NLS) models
predict wave-breaking phenomena in the limit of vanishing dispersion. Such a condition, for temporal shocks, is in general very
difficult to achieve over wide spectral regions. PCFs can however be designed with almost flat dispersion over bandwidths of several
hundreds nanometers, thus opening up new perspectives in temporal optical shock generation and related applications.

From a theoretical point of view, there are still several open problems concerning the physics of optical shocks. One of the
most challenging is the onset of wave-breaking phenomena in the regime of focusing nonlinearity and anomalous group velocity dispersion.
Even if unstable kink-antikink solutions (regularized shock fronts without oscillations) are known to exist 
when including Raman terms  \cite{Agrawal92,Kivshar93}, the standard hydrodynamical limit of the NLS apparently rules out shock formation in the focusing regime, 
as the jump condition on the resulting Euler equations (see, e.g, \cite{hoefer06})  gives an imaginary result,
which does not have a simple physical interpretation. 
However, it has been shown that, in non-local models \cite{ghofraniha07} for spatial beam propagation and BEC, 
a multi-valued solution of the hydrodynamical equation for the so-called velocity field can be predicted by using the method of characteristics. 
Such a solution corresponds in the numerical simulations of the nonlocal NLS to the clear formation
of an undular-like bore in a regime in which it competes with the modulational instability, as also considered in \cite{Assanto08shock}.
The scenario is hence extremely rich and interesting, and the open question is whether temporal shocks, undular bores and wave-breaking phenomena
may form in a focusing regime when considering a real-world system.

The most natural counterpart of a nonlocal nonlinear response in the time domain is the Raman effect. Indeed this is described
by a kernel response function, which, under suitable conditions on the pulse duration, leads to a nonlinear shock term containing
the first-order derivative of the intensity \cite{AgrawalBook}.
Previous investigations of intense light propagation in solid-core PCFs when including the Raman effect have shown that the peculiar breathing phenomenon exhibited by higher-order solitons can be used to excite the formation of quasi-symmetric resonant radiation in a step-like fashion in presence of highly distorted GVD \cite{biancalana09}. However, in that work, and in all previous observations of the phenomenon (see e.g. \cite{cristiani}), there is no convincing explanation of why the soliton breathing should increase its rate in the presence of the Raman effect, which leads to soliton splitting, and the internal dynamics of the breathing has currently a rather cumbersome interpretation.

In this paper we show that under suitable conditions the soliton breathing process is accompanied by the formation
of multiple optical shock waves leading to strong spectral broadening, and that this can described by a hydrodynamical model containing a gravity-like slope term, or equivalently a constant external electric field in a cold plasma. 
This point of view allows one to clarify and shed new light on the several well-known phenomena described above.
Our theoretical results are validated by real-world simulations in a realistic PCF, and represent the first theoretical prediction of shock waves in the focusing regime, also unveiling their measurable signatures in the XFROG signal.

The paper is organized as follows. In section \ref{hydrodynamical} we discuss the hydrodynamical model for the NLS in the presence of a Raman term.
In section \ref{particleinterpretation} we adopt the method of characteristics to predict and give a mechanical interpretation to the occurrence
of multi-valued solutions for the velocity-field. In section \ref{simulations} we describe the PCF we used for investigating
temporal hydrodynamic-like shocks and report on the corresponding simulations and the XFROG signal at the shock formation. 
Conclusions are drawn in section \ref{conclusions}.

\section{The hydrodynamical limit in the accelerated reference frame} \label{hydrodynamical}
We start by considering the simplest model for an optical pulse propagating in an optical fibre, described by the envelope equation \cite{AgrawalBook}
\begin{equation}
i \psi_z+\frac{1}{2}\psi_{tt}+|\psi|^2\psi -\tau_R \psi \partial_t |\psi|^2=0.
\label{NLS1}
\end{equation}
In Eq. (\ref{NLS1}), $\psi$ is the envelope of the electric field, scaled with the soliton power $P_{0}=|\beta_{2}|/(\gamma t_{0})$, where $\beta_{2}$ is the second order dispersion coefficient, $\gamma$ is the nonlinear coefficient of the fiber, and $t_{0}$ is the input pulse duration. The dimensionless propagation length $z$ is scaled with the second order dispersion length $L_{\rm D2}=t_{0}^{2}/|\beta_{2}|$, while the temporal variable $t$ is scaled with $t_{0}$. The last term in (\ref{NLS1}) represents the Raman effect, and $\tau_{R}\equiv T_{R}/t_{0}$, where $T_{R}$ is the Raman response time (about $2$ fs in silica) \cite{AgrawalBook}. For simplicity, only second order dispersion has been included in the model of Eq. (\ref{NLS1}). Note that Eq. (\ref{NLS1}) is written in the anomalous dispersion regime, where bright solitons are expected in the absence of the Raman term ($\tau_R=0$). Eq. (\ref{NLS1}) is subject to the initial condition
\begin{equation}
\psi(z=0,t)=N\sech{t}. \label{initialcond}
\end{equation}
The generation of the shock-like dynamics corresponds to the case of large $N$, and introducing the smallness parameter
$\epsilon\sim1/\sqrt{N}$, the hydrodynamical limit is obtained via rewriting Eq. (\ref{NLS1}) in the semiclassical form with the scaled variables
$z\rightarrow z/\epsilon^3 $, $t\rightarrow t/\epsilon^2 $, $\psi\rightarrow \psi \epsilon$, obtaining
\begin{equation}
i\epsilon \psi_z+\frac{\epsilon^2}{2}\psi_{tt}+|\psi|^2\psi -\frac{\tau_R}{\epsilon} \psi \partial_t |\psi|^2=0\text{.}
\label{NLS2}
\end{equation}
The Raman term appears at the relevant order in the hydrodynamical limit for $\tau\sim \epsilon$, 
which is the condition at which the Raman effect radically alters the shock dynamics, as also confirmed by the simulations 
reported below.

It is interesting to derive the hydrodynamical limit in the accelerated reference
frame where the soliton moves in presence of the Raman effect 
\cite{skryabin1}. This is obtained by introducing the variable 
$\xi=t-\frac{g}{2} z^2$ (with $g=32\tau_{R}a^{2}/15$ as in \cite{skryabin1}), and performing the Gagnon-B\'elanger phase transformation \cite{gagnon}
\begin{equation}
 \psi=f(\xi,z)\exp[ -i (\frac{g^2}{3} z^3 - g z t + a z)],
\end{equation}
after which the resulting NLS is written as \cite{akhmediev,gagnon}
\begin{equation}
i f_z- g \xi f +  a f +\frac{1}{2}f_{\xi\xi} +|f|^2 f -\tau_R f \partial_\xi |f|^2=0
\label{NLS3}
\end{equation}
Letting $f=\sqrt{\rho} \exp(i\phi)$,
and performing a WKB expansion on Eq. (\ref{NLS3}), one obtains
\begin{eqnarray}
  \rho_z+\partial_\xi(\rho v)=0\label{rhoxi}\\
  \phi_z+\frac{1}{2}\phi_\xi^2=- g \xi + a+\rho-\tau_R \partial_\xi \rho
+\frac{1}{4}\frac{1}{\sqrt{\rho}}\frac{\partial}{\partial \xi}\frac{\rho_\xi}{\sqrt{\rho}}
 \label{rhophi}
\end{eqnarray}
By deriving Eq. (\ref{rhophi}) with respect to $\xi$, one has (after defining the {\em velocity field} $v=\phi_\xi$, physically corresponding to the instantaneous frequency inside the pulse)
\begin{equation}
 v_z+ v v_\xi=-\partial_\xi (U_{QP}+U).
\label{vacc}
\end{equation}
The {\em quantum pressure potential} is defined as
\begin{equation}
 U_{QP}=-\frac{1}{4}\frac{1}{\sqrt{\rho}}\frac{\partial}{\partial_\xi}\frac{\rho_\xi}{\sqrt{\rho}},
 \label{quantumpressure}
\end{equation}
and the modified 'classical' potential term is
\begin{equation}
 U=g \xi - \rho +\tau_R \partial_\xi \rho.
\label{hydropotacc}
\end{equation}
Note that potential $U$ in Eq. (\ref{hydropotacc}) is exactly the same as the one formerly introduced in \cite{skryabin1}.
The above relations show that the hydrodynamical limit of the NLS with Raman corresponds to
fluid motion in an accelerated frame (being $g$ the acceleration) or, equivalently, to a charged plasma in a constant electric field of magnitude $g$.

\section{Effective particle interpretation} \label{particleinterpretation}
Eq. (\ref{vacc}) can be solved before the occurrence of shocks in an approximate way by recalling that
the equation for $\rho$ is obtained at a higher order in $\eps$, and hence that the initial dynamics
is ruled by the velocity field $v$ only. This allows, as a zero-order approximation, to
take $\rho(\xi)$ as a fixed profile, e.g. as a sech function denoting the
initial pulse profile. Eq. (\ref{vacc}) can hence be mapped by the method of characteristics into the system of
ordinary differential equations
\begin{equation}
 \begin{array}{l}
  \frac{d\xi}{dz}=v\\
  \frac{d v}{dz}=-\partial_\xi U(\xi)
 \end{array}
\label{odeshock}
\end{equation}
with the initial conditions $\xi=s$ and $v=0$ at $z=0$.
The solution for a given $\rho(\xi)$, gives the manifold $(\xi(s,z),v(s,z))$, which for a given
$z$ provides a parametric plot of $v$ versus $z$ and allows one to predict the occurrence of hydrodynamical shocks.

Eqs. (\ref{odeshock}) can be interpreted as the motion of a particle with trajectory
$\xi(z)$ in the potential $U$ and can be rewritten as a single Newton-like equation
\begin{equation}
 \frac{d^2 \xi}{d z^2}=-\frac{\partial U}{\partial \xi}.
\end{equation}
Shocks correspond to the case in which the plot of the trajectory in the
space $(\xi,v)$ display a vertical slope, i.e. $d\xi/dv=0$.

\begin{figure}
\includegraphics[width=8.3cm]{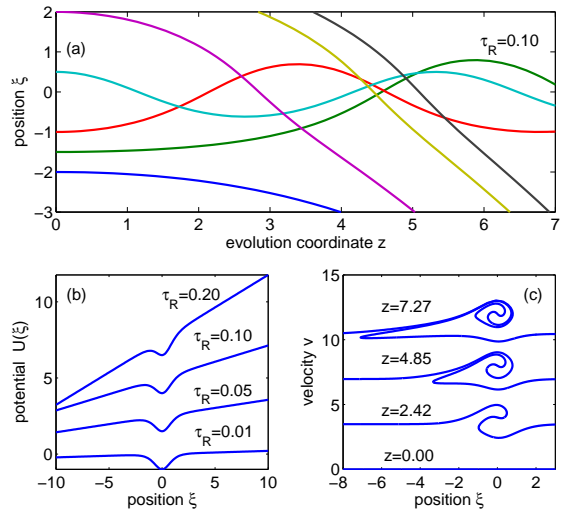}
 \caption{(Color online)
 (a) Positions of a particle falling inside potential $U(\xi)$ as described by Eq. (\ref{hydropotacc}), for different values of the initial position $\xi=s$ and for $\tau_{R}=0.1$. Multiple collisions between oscillating bound particles produce multiple values for the velocity field $v$, i.e. shocks. Non-bound particles can also collide with bound particles -- although more rarely -- contributing to isolated shocks.  (b) Shape of the effective potential for different values of $\tau_R$ with $\rho=\sech{\xi}^2$; (c)
Solution of the hydrodynamical system (\ref{odeshock}) for different values of $z$, showing multiple-shock occurrences.
The velocity is plotted for clarity with an arbitrary vertical shift for the various $z$.
\label{fighydro1}}
\end{figure}

This can be geometrically interpreted as follows: various trajectories are generated by varying the particle's initial position
$\xi=s$, see Fig. \ref{fighydro1}(a), which corresponds to considering the motion of particles in the potential $U$ [shown in Fig. \ref{fighydro1}(b)] for different values of $\tau_R$ which are falling (with initially zero velocity) from position $\xi=s$. A shock, i.e. a multi-valued function $v=v(\xi)$ 
in the plane $(\xi,v)$ [see Fig. \ref{fighydro1}(c)] corresponds to the existence of trajectories reaching the same position 
$\xi$ at the same $z$ with different velocities, i.e. to {\em collisions between particles} falling from different initial positions under the effect of $U$.

Given the shape of $U(\xi)$, such collisions specifically involve the particles trapped in a bounded motion,
which implies the formation of shocks in a periodic fashion, see also Fig. \ref{fighydro1}(a) and (c).

As $\tau_R$ increases the bounded motion occurs within an increasingly smaller region in $\xi$ [the potential well is reduced, as it is clear from Fig. \ref{fighydro1}(b)], and it becomes negligible for larger values of $\tau_{R}$. This implies that for increasing $\tau_R$, shocks become more frequent (as the corresponding particle collisions), but also that the corresponding discontinuities will be less pronounced.
Such a qualitative interpretation, gives a simple picture of what is observed in the simulations discussed below.

\section{Simulations}\label{simulations}
In this section we show that the hydrodynamical limit of the NLS equation [see Eq. (\ref{NLS2})] can be realized in practice through a realistic PCF design. We also compare the simulation dynamics of the idealized model of Eq. (\ref{NLS2}) with a more realistic simulation based on the full generalized NLS equation (GNLSE).

\begin{figure}
\includegraphics[width=6cm]{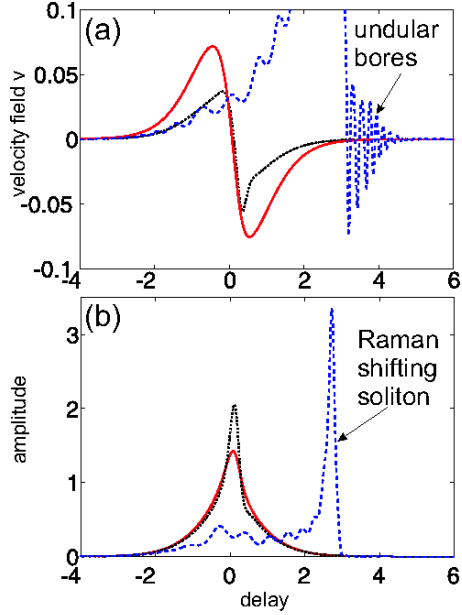}
 \caption{(Color online) Pulse propagation and shock formation according to Eq. (\ref{gnlse1}), using the GVD of Fig. \ref{figgvd}.  (a) Velocity field $v=\de_{t}\phi$ as a function of time for three different propagation distances, namely $z=0.4$ (red solid line, before the occurrence of shock), $z=0.62$ (black dotted line, at the moment of shock) and $z=1.5$ (blue dashed line, long distance dynamics), for $\tau_{R}=0.1$ and $\eps=0.2$. For long propagation dynamics, the formation of clear undular bores is observed. (b) Amplitude $|\psi|$ at the same values of $z$ as in (a). At the moment of the first shock, a sharp cusp is formed. \label{nlsshock}}
\end{figure}

We start by integrating numerically Eq. (\ref{NLS2}) with parameters $\tau_{R}=0.1$ and $\eps=0.2$, with initial condition $\psi_{\rm in}=\sech{t}$. This set of parameters is chosen to realize the hydrodynamical limit by means of the scaling discussed in section \ref{hydrodynamical}. After a propagation length of $z\simeq 0.62$, one can observe the formation of the first shock, see Fig. \ref{nlsshock}(a) where the time-derivative of the phase (i.e. the velocity field) $v(t)\equiv\phi_{t}$ has been plotted. Linked to the formation of the phase shock, one observes the formation of a sharp cusp in the time domain located at the maximum amplitude of the pulse, see Fig. \ref{nlsshock}(b). Large values of the time-derivative at the cusp produce large spectral broadening in the frequency domain, which in the conventional and very well-known picture is associated with the temporal and spectral 'breathing' oscillations of higher-order solitons \cite{biancalana09,AgrawalBook}. In the presence of the Raman effect ($\tau_{R}\neq 0$), one will observe the occurrence of multiple shocks, which become more and more frequent during propagation, contrary to the perfectly regular oscillations of the soliton breathing phenomenon in the absence of the Raman \cite{biancalana09,AgrawalBook}. This 'breathing acceleration' can again be interpreted as multiple shocks of the falling fluid induced by the 'gravity-like' potential of Eq. (\ref{vacc}), as described in section \ref{particleinterpretation}. Moreover, as is shown in Fig. \ref{nlsshock}(a), a very clear undular bore phenomenon develops in the trailing edge of the Raman-shifting pulse at longer propagation distances, which does not occur in the absence of the Raman effect. This is associated to the strong oscillations of the velocity field $v$ induced by the quantum pressure potential (\ref{quantumpressure}). This term becomes indeed important in proximity of those parts of the pulse for which $\rho\rightarrow 0$, i.e. far from the pulse center. This, when combined with the Raman gravity-like potential (\ref{hydropotacc}), leads to the formation and transport of increasingly stronger oscillations in the trailing edge of each pulse.

\begin{figure}
\includegraphics[width=7cm]{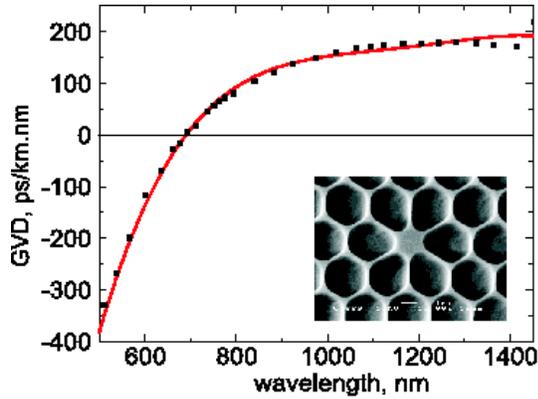}
 \caption{(Color online) GVD curve of the PCF used in our simulations. The black dots represent the data measured experimentally through white-light interferometry, while the solid curve is a polynomial fit. The ZDW is close to $\lambda_{Z}=690$ nm. The inset shows an SEM of the fiber cross section.
\label{figgvd}}
\end{figure}

We now compare the shock dynamics of the above simplified model with the full GNLSE, applied to a realistic waveguide. In Fig. \ref{figgvd} we show the GVD of a solid-core PCF, the SEM picture of which is given in the inset. The black dots correspond to measured data, while the red solid line shows a fit. The fiber has a core diameter of approximately $1.5$ $\mu$m, an estimated nonlinear coefficient $\gamma\simeq 100$ W$^{-1}$m$^{-1}$, and a zero-GVD point located at $\lambda_{Z}\simeq 690$ nm. We use the following GNLSE \cite{AgrawalBook}: \eq{gnlse1}{\small{i\de_{z}\psi+\hat{D}(i\de_{t})\psi+\left(1+\frac{i\de_{t}}{\omega_{0}t_{0}}\right)\left[\psi\int_{-\infty}^{+\infty}r(t-t')|\psi(t')|^{2}dt'\right]=0},} where $r(t)=[(1-\theta)\delta(t)+\theta h(t)]/t_{0}$ is the total response function, $t_{0}$ is the input pulse duration, $\theta$ is the relative contribution between the non-instantaneous Raman and the instantaneous Kerr effect in the material ($\theta=0.18$ in silica), $\delta(t)$ is the Dirac-delta function. The Raman response function is introduced in the code by using the Hollenbeck and Cantrell approach \cite{ramandata}. The $h$-function is normalized, $\int_{-\infty}^{+\infty}r(t')dt'=1$, while the dimensionless Raman response time parameter used in Eq. (\ref{NLS1}) and Eq. (\ref{NLS2}), proportional to the first momentum, is given by $\tau_{R}=(1/t_{0})\int_{-\infty}^{+\infty}t'r(t')dt'$. Operator $\hat{D}(i\de_{t})$ in Eq. (\ref{gnlse1}) describes the full complexity of the fiber GVD shown in Fig. \ref{figgvd}, and is given by $\hat{D}(i\de_{t})\equiv\sum_{m\geq 2}\frac{\beta_{m}(i\de_{t})^{m}}{m!}$, where $\beta_{m}\equiv(\de^{m}\beta(\omega)/\de\omega^{m})_{\omega=\omega_{0}}$ is the $m$-th derivative of the propagation constant $\beta(\omega)$ calculated at a suitable arbitrary reference frequency $\omega_{0}$.
\begin{figure}
\includegraphics[width=8.3cm]{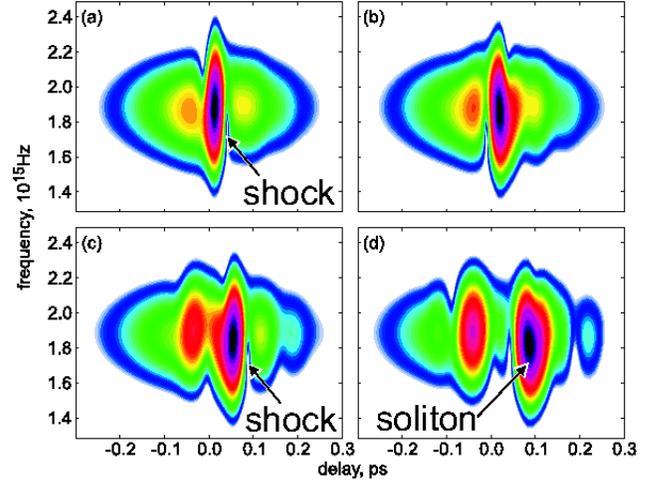}
\caption{(Color online) XFROG spectrograms for the pulses at $z$ equal to (a) $1.761$, (b) $2.192$, (c) $2.723$, and (d) $3.393$ cm. The peak power of the $t_{0}=130$ fs input pulse was $P=5$ kW ($N=3.1$), and they were launched at $\lambda=1$ $\mu$m.
Fig (a) and (b) are snapshots of the first shock. Fig (c) and (d) show the second shock, coinciding with the second spectral oscillation. \label{figxfrog}}
\end{figure}

The panels in Fig. \ref{figxfrog}(a-d) show the evolution of the XFROG trace $I(\delta,\tau)=\left| A(t)A_{\rm ref}(t-\tau)e^{-i\delta t}dt \right|^{2}$ \cite{AgrawalBook} ($A_{\rm ref}$ is a suitable reference pulse) recorded at four different propagation distances, corresponding to the first two occurrences of maximum spectral broadening, for $\lambda=1$ $\mu$m, relatively far from the zero GVD point.

\begin{figure}
\includegraphics[width=10cm]{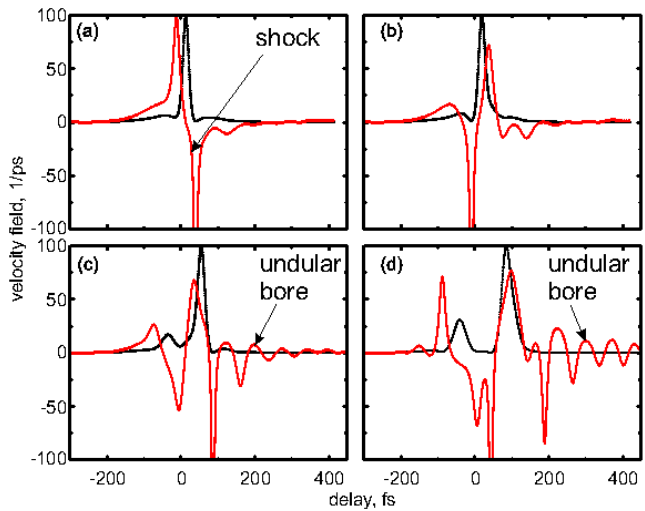}
\caption{(Color online) (a-d) Amplitude (black dotted line) and velocity field (red solid line) in the same conditions as in Figs. \ref{figxfrog}(a-d). (a) is the moment of the first shock, while (b-d) show the generation of undular bores from the strong velocity field oscillations of the singularities due to the quantum pressure potential (\ref{quantumpressure}). \label{figpanel}}
\end{figure}

The goal of our simulations is to identify the multiple shock generation during pulse propagation in a PCF. 
In section II we showed that the hydrodynamical limit requires large N, thus very high peak powers.
We find however, that in our siumlations the dynamics of the multiple shock can best be demonstrated for lower N.
In this case, the dynamics are not obscured by unwanted effects like resonant radiation, which disturb the 
soliton propagation and prevents thus a clear signature of the multiple shocks. In this regard, the presented GVD curve in Fig. \ref{figgvd} will be advanteous for the purpose of observing shock phenomena. A higher value of $\beta_{2}$ would lead to a massive Raman self-frequency shift which would obscure shock formation. On the other hand, smaller values of $\beta_{2}$ (or too large values of $N$) lead to a dominating self-phase modulation, since $L_{\rm D2}$ would be much longer that the nonlinear length $L_{\rm NL}=(\gamma P)^{-1}$. In that case, a very strong spectral broadening would be accompanied by the emission of resonant radiation from solitons \cite{biancalanaresonant}, which would also disturb the formation of clear shocks.
The fiber is pumped by a sech pulse with $N=3.1$. The amplitude shocks that develop at these moments are evident. The panels in Fig. \ref{figpanel}(a-d) show the corresponding intensity profiles and the velocity field $v(t)=\de_{t}\phi$ at the same values of $z$ as in Fig. \ref{figxfrog}. Again, one can see phase discontinuities located at the various pulse centers. Moreover, one can notice the generation of strong undular bores in the trailing edge of the pulse, which agree qualitatively with the simplified model examined above. 

\begin{figure}
\includegraphics[width=8.3cm]{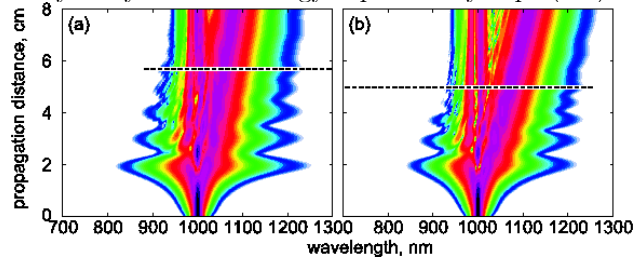}
 \caption{(Color online) Spectral evolution of $P=5$ kW pulses for Raman fractions (a) $f_{R}=0.18$ ($\tau_{R}=0.0334$), and (b) $f_{R}=0.3$ ($\tau_{R}=0.0557$), with $t_{0}=130$ fs and a pulse launched at $\lambda=800$ nm in the PCF of Fig. \ref{figgvd}. For increasing values of $\tau_{R}$, the spectral oscillation rate rises, which can be interpreted by the hydrodynamical formulation of  Eqs. (\ref{rhoxi}-\ref{rhophi}). Horizontal dashed lines indicate the distance at which one has 5 spectral oscillations, to help visualization. \label{figbreathing}}
\end{figure}
In Figs. \ref{figbreathing}(a,b) we show the evolution of Eq. (\ref{gnlse1}) in two different cases, one for $\tau_{R}=0.0334$ [Fig. \ref{figbreathing}(a)] and the other for $\tau_{R}=0.0557$ [Fig. \ref{figbreathing}(b)]. One can notice the well-known slight increase of the rate of spectral oscillations when increasing the value of $\tau_{R}$, which 
cannot find an easy explanation in the model of Eq. (\ref{gnlse1}). However, such behavior can easily be explained in terms of the hydrodynamical analogy expressed by Eqs. (\ref{rhoxi}-\ref{rhophi}): in presence of Raman effect, the 'falling' photon fluid feels a gravity-like field which increases the rate of shocks during propagation, according to the qualitative explanation based on colliding particles given in section \ref{particleinterpretation}.

\section{Conclusions}\label{conclusions}
In conclusion we have shown that suitably 
engineered photonic crystal fibers may sustain the 
formation of multiple hydrodynamical-like shocks during the
propagation of ultrashort pulses.
The shocks are shown to be clearly evident in the XFROG signal,
have measurable signatures, and occur for 
a focusing nonlinearity. This effect can be exploited for the optimization
and control of supercontinuum generation, and is expected
to play a role when considering multidimensional dynamics.

\section{Acknowledgements}
We acknowledge support from the INFM-CINECA initiative
for parallel computing and CASPUR.
The research leading to these results has received
funding from the European Research Council under the European Community
Seventh Framework Program (FP7/2007-2013)/ERC grant agreement n.201766. FB, SS and PSJR are supported by the Max Planck Society for the Advancement of Science (MPG).

\end{document}